\documentclass[12pt]{article}
\usepackage[hypertex]{hyperref}
\usepackage[dvipdfmx]{graphicx} 
\usepackage{graphicx,amsmath,amssymb,amsfonts,cite,bm,cancel}

 \def\b{\beta}    \def\D{\Delta}   \def\h{\eta} \def\th{\theta}   \def\l{\lambda}  \def\m{\mu} \def\n{\nu}     \def\r{\rho} \def\s{\sigma}   \def\ph{\phi} \def\Ph{\Phi}     

\def\dg{\dagger}

\newcommand{\Lg}{\mathcal{L}}

\newcommand{\Gets}{\Leftarrow}

\newcommand{\Getsto}{\Leftrightarrow}

\newcommand{\vev}[1]{ \langle {#1} \rangle }

\newcommand{\tr}{{\rm tr}}

\newcommand{\row}[2]{ \begin{pmatrix}  #1 & #2   \end{pmatrix}  }
\newcommand{\column}[2]{ \begin{pmatrix}  #1 \\ #2 \\  \end{pmatrix} }
\newcommand{\diag}[2]{ \begin{pmatrix}  #1 & 0 \\ 0 & #2 \\   \end{pmatrix}  }
\newcommand{\offdiag}[2]{ \begin{pmatrix} 0 & #1 \\ #2 & 0 \\   \end{pmatrix} }

\newcommand{\RRow}[4]{ \begin{pmatrix} #1 & #2 & #3 & #4 \end{pmatrix} }
\newcommand{\CColumn}[4]{ \begin{pmatrix} #1 \\ #2 \\ #3 \\ #4 \end{pmatrix} }

\newcommand{\Rotleft}[1]{ 
\begin{pmatrix}
\cos {#1} & \sin {#1} \\
- \sin {#1} & \cos {#1} \\
\end{pmatrix}
}
\newcommand{\Rotright}[1]{ 
\begin{pmatrix}
\cos {#1} & -\sin {#1} \\
\sin {#1} & \cos {#1} \\
\end{pmatrix}
}

\def\M{{\mathcal M}}

\begin{document}

%\begin{titlepage}

\begin{flushright}
STUPP-18-234
\end{flushright}

\vskip 1.35cm

\begin{center}
{\Large \bf Flavor structure from flavored Higgs mixing}

\vskip 1.2cm

Masaki J. S. Yang

\vskip 0.4cm

{\it Department of Physics, Saitama University, \\
Shimo-okubo, Sakura-ku, Saitama, 338-8570, Japan\\
}
%\date{\today}

%%%%%%%%%%%%%%%%%%%%%%%%%%%%%
\begin{abstract} %%%%%%%%%%%%%%%%%%%%%
%%%%%%%%%%%%%%%%%%%%%%%%%%%%%

In this paper, we suggest a simple model 
which induces realistic flavor structure from 
mixing of flavored Higgs doublets. 
The idea is based on the decoupling limit. 
In a model with many Higgs doublets, 
 the mass eigenstates of scalars are linear combinations of Higgs doublets. 
If the mass matrix of Higgs fields has only one massless mode, 
and if the linear combination has flavor dependence, 
the induced Yukawa coupling will have nontrivial flavor structure.

We suggest a mass matrix of flavored Higgs fields in a $U(2)_{L} \times U(2)_{R}$ toy model. 
An advantage of this model 
is that all of the elements in Yukawa matrix 
can be determined from renormalizable Higgs potential. 

%%%%%%%%%%%%%%%%%%%%%%%%%%%%%
\end{abstract} %%%%%%%%%%%%%%%%%%%%%%
%%%%%%%%%%%%%%%%%%%%%%%%%%%%%

\end{center}
%\end{titlepage}

%%%%%%%%%%%%%%%%%%%%
\section{Introduction}
%%%%%%%%%%%%%%%%%%%%

The discovery of the Higgs boson  \cite{Aad:2012tfa,Chatrchyan:2012ufa} 
and its coupling analysis \cite{Khachatryan:2014jba} 
reinforce the experimental validity of the Standard Model (SM).
However, the model still has several mysteries, 
{\it e.g.,} theoretical origin of Higgs, the hierarchy problem, and the flavor puzzle.

If some flavored Higgs field exists \cite{Koide:1995xk,DiazCruz:2002er, Ko:2011di, Barradas-Guevara:2014yoa},  
flavor structure can be mediated to the Higgs sector 
by radiative or mixing mechanisms \cite{DiazCruz:2002er}.
In the mixing context, 
several ideas induce flavor structures from mixing between SM fermions and heavy fermionic partners, 
such as universal seesaw mechanism \cite{Berezhiani:1983hm,Berezhiani:1985in}, partial compositeness \cite{Kaplan:1991dc, Contino:2006nn} and $E_{6}$ twist mechanism \cite{Bando:2000gs,Maekawa:2001uk}.
However, the flavor structure perhaps originates in the scalar mixing. 
This idea cannot be found as far as author knows.

In this paper, we suggest a simple model 
which induces realistic flavor structure from 
mixing of flavored Higgs doublets. 
The idea is based on the decoupling limit \cite{Haber:1994mt,Gunion:2002zf}. 
In a model with many Higgs doublets ($n$HDM), 
 the mass eigenstates of scalars are linear combinations of Higgs doublets. 
If the mass matrix of Higgs fields has only one massless mode, 
and if the linear combination has flavor dependence, 
the induced Yukawa coupling will have nontrivial flavor structure.
We suggest a mass matrix of flavored Higgs fields in a $U(2)_{L} \times U(2)_{R}$ toy model. 

An advantage of this model 
is that all of the elements in Yukawa matrix 
can be determined from renormalizable Higgs potential. 
This feature is substantially different from Froggatt--Nielsen mechanism \cite{Froggatt:1978nt}  
with many undetermined $\mathcal O(1)$ couplings. 
A drawbacks are arbitrariness of mass parameters 
and fine-tunings. The heavy flavored Higgs bosons have ``natural mass''. 
By contrast, the mass of the lightest SM Higgs boson should be fine-tuned. 
If the flavored Higgs fields are gauged Nambu--Goldstone bosons (NGBs), 
such as in composite Higgs \cite{Georgi:1984af,Kaplan:1983fs,Agashe:2004rs}, Little Higgs \cite{ArkaniHamed:2002qy,ArkaniHamed:2002qx}, twin Higgs \cite{Chacko:2005pe}, 
GIFT mechanism with supersymmetry \cite{Inoue:1985cw}, 
these theories require TeV-scale partner fields that have strong tension 
with current experiments. 
Therefore, in this idea, it is difficult to avoid fine-tuning of the lightest Higgs mass.

Meanwhile, if the determinant of the mass matrix $\det \M < 0$ holds, mass of the lighter boson becomes negative and 
triggers the electroweak symmetry breaking
(bosonic seesaw mechanism) \cite{Calmet:2002rf, Haba:2015lka,Haba:2015yfa}. 
Interplay between these ideas also seems to be interesting possibility.

%%%%%%%%%%%%%%%%%%%%
\section{Basic idea: the decoupling limit}
%%%%%%%%%%%%%%%%%%%%

The basic idea of this paper is based on the ``decoupling limit'' \cite{Haber:1994mt,Gunion:2002zf}. 
In the two Higgs doublet model (2HDM), the mass eigenstates of Higgs fields  are linear combinations of two Higgs doublets 
$\ph \sim a \ph_{1} + b \ph_{2}$. 
If one of the mass eigenstates have enough large mass, 
the heavy state will be integrated out from the theory 
and overall factor of the Yukawa matrices will be changed. 
In this section, we review this mechanism. 

For the Higgs doublets $\ph_{1}, \ph_{2}$, 
the Higgs potential of the 2HDM is written by \cite{Branco:2011iw}
\begin{align}
V_{H} (\ph_{1}, \ph_{2})  = 
\sum_{a,b}^{1,2} m_{ab} \phi^{\dg}_{a} \phi_{b} + 
\sum_{a,b,c,d}^{1,2} \l_{ab , cd} (\phi^{\dg}_{a} \phi_{b}) (\phi^{\dg}_{c} \phi_{d}) .
\end{align}
The mass matrix $\M$ is given by
\begin{align}
\ph^{\dg}_{a} \M_{ab} \ph_{b} \equiv 
\row{\ph_{1}^{\dg}}{\ph_{2}^{\dg}}
\begin{pmatrix}
m_{11}^{2} & m_{11} m_{22} \\ 
m_{11} m_{22} & m_{22}^{2}
\end{pmatrix} 
\column{\ph_{1}}{\ph_{2}} . 
%\to \infty , ~~ 
\end{align}
Here, we assumed reality condition $m_{21}^{2} = m_{12}^{2 *}$ and 
 $\det \M = m_{11}^{2} m_{22}^{2} - m_{12}^{4} = 0. $
This matrix is diagonalized as
\begin{align}
\begin{pmatrix}
m_{1}^{2} & m_{12}^{2} \\
m_{12}^{2} & m_{2}^{2}
\end{pmatrix}
=
\Rotright{\b} 
\diag{0}{m_{11}^{2} + m_{22}^{2}}
\Rotleft{\b} .
\end{align}
The mass eigenstates $\ph, \ph_{H}$ 
are found to be
\begin{align}
 \column{\ph}{\ph_{H}} 
= 
\Rotleft{\b} \column{\ph_{1}}{\ph_{2}}  
~~ \Getsto ~~ 
\column{\ph_{1}}{\ph_{2}}
= 
\Rotright{\b}
 \column{\ph}{\ph_{H}} . 
 \label{rotbeta}
\end{align}
Then the massive eigenstate $\ph_{H}$ decouples from the low-energy theory 
(decoupling limit). 
If the full potential of $\ph_{i}$ has symmetry breaking minima and $\ph_{1,2}$ develop vevs $v_{1,2}$, 
the decoupling is realized as far as $m_{ij} \gg v_{i} (i,j = 1,2)$ holds.

Yukawa interactions of quarks (with natural flavor conservation \cite{Glashow:1976nt}) are
\begin{align}
\Lg_{Y} = Y_{u} \bar q_{L} u_{R} \tilde \ph_{1} + 
Y_{d} \bar q_{L} d_{R} \ph_{2} + h.c. . 
\end{align}
When the massive state $\ph_{H}$ is integrated out, 
it leads to 
\begin{align}
\Lg_{Y} &= Y_{u} \bar q_{L} u_{R}  (c_{\b} \tilde \ph - \cancel{s_{\b} \tilde \ph_{H}}) + 
Y_{d} \bar q_{L} d_{R} (s_{\b} \ph + \cancel{c_{\b} \ph_{H}}) + h.c.  \\
& \to c_{\b}  Y_{u} \bar q_{L} u_{R} \tilde \ph  + 
s_{\b} Y_{d} \bar q_{L} d_{R}  \ph+ h.c. , 
\end{align}
and the isospin violation of Yukawa interactions 
is introduced naturally. 
In some sense, the decoupling limit of 2HDM induces 
the simplest flavor structure. 
Therefore, if we extend this idea to the flavor space, 
the nontrivial flavor structure will be induced 
from flavored Higgs mixing. 

%%%%%%%%%%%%%%%%%%%%%%%%
\section{$U(2)_{L} \times U(2)_{R}$ toy model}
%%%%%%%%%%%%%%%%%%%%%%%%

For simplicity, 
we consider a toy model with $U(2)_{L} \times U(2)_{R}$ flavor symmetry. 
The Higgs doublets are enlarged to 
a $U(2)_{L} \times U(2)_{R}$ bi-doublet scalar $\ph_{ij}^{ab}$ 
with gauge and flavor indices $a,b,i,j = 1,2$. 
%Then it is an eight Higgs doublet model (8HDM). 
Left-(right-)handed quarks are assigned to doublets of $U(2)_{L (R)}$. 
There is an additional scalar field $\h_{ij}$ that breaks the flavor symmetry.
Table 1 shows the charge assignments of the fermions and Higgs fields under the gauge and the flavor symmetries.
\begin{table}[htb]
  \begin{center}
    \begin{tabular}{|c|cc|cc|} \hline
           & $SU(2)_{L}$ & $SU(2)_{R}$ & $U(2)_{L}$ & $U(2)_{R}$ \\ \hline \hline
      $q_{Li}^{a}$  & \bf 2 & \bf 1 & $\bf 2_L$ & $\bf 1_R$    \\
      $q_{Ri}^{a}$  & \bf 1 & \bf 2 & $\bf 1_L$ &  $\bf 2_R$ \\ \hline
      $\phi_{ij}^{a b}$  & \bf 2 & \bf 2 & $\bf 2_L$ &  $\bf 2_{R}$ \\ 
      $\h_{ij}$  & \bf 1 & \bf 1 & $\bf 2_L$ &  $\bf 2_{R}$ \\ \hline
%      $\D_{R}$  & \bf 1 & \bf 3 & $\bf 1_L$ &  $\bf 1_R $ \\ \hline
    \end{tabular}
    \caption{The charge assignments of the fermions and Higgs fields under the gauge and the flavor symmetries.}
  \end{center}
\end{table}

\vspace{-18pt}

For these fields, the Yukawa interactions are given by
\begin{align}
\Lg_{Y} = 
Y \bar q_{Li}^{a} \ph_{ij}^{ab} q^{b}_{Rj} + h.c. . 
\end{align}
Here, $Y$ is just a coupling and does not have any flavor structures. 
The coupling between fermions and the scalar $\h_{ij}$ 
is forbidden by the gauge symmetry. 

The general (flavor breaking) mass matrix of $\ph_{ij}$ is written as
\begin{align}
\ph^{\dg}_{\m} \M_{\m\n} \ph_{\n} \equiv 
\row{\ph_{0}^{\dg}}{\ph_{i}^{\dg}}
\begin{pmatrix}
m_{00}^{2} & m_{0j}^{2} \\
m_{i0}^{2} & m_{ij}^{2}
\end{pmatrix}
\column{\ph_{0}}{\ph_{j}} .
\end{align}
Here, we rewrite the bi-doublet Higgs as 
\begin{align}
\phi_{ij} = \s^{\m}_{ij} \phi_{\m} , ~~~ 
\ph_{\n} = {1 \over 2} \tr [ \s^{\n} \ph_{ij}] , 
~~ \s^{\m} \equiv (1_{2}, \s^{1}, \s^{2}, \s^{3}).
\end{align}
If $\ph_{ij}$ is a hermitian matrix, $\ph_{\n}$ becomes real field. 
In this case, $\ph_{ij}$ is a general complex matrix and then
$\ph_{\n}$ is also complex field. 

The mass matrix is diagonalized by a unitary matrix $U$:
\begin{align}
\ph_{\m}^{\dg} \M_{\m\n} \ph_{\n} 
= \ph_{\m}^{\dg} U_{\m \r} \M_{\r}^{\rm diag} U_{\r \n}^{\dg} \ph_{\n} .
\end{align}
Then, the mass eigenstates $\Ph_{\r}$ are found to be 
 $\Ph_{\r} = U_{\r \n}^{\dg} \ph_{\n}$ (and then $\ph_{\m} = U_{\m \n} \Ph_{\n}$).
Here we assume the eigenvalues of $\M$ has only one zero, such as
$\M_{\r}^{\rm diag} \sim {\rm diag}(0,M,M,M)$. 
Then $\Ph_{0}$ is the massless mode and massive modes $\Ph_{i}$ will be integrated out. 
The Yukawa interactions are modified as 
\begin{align}
\Lg_{Y} = Y \bar q_{Li} \ph_{ij} q_{Rj} 
%& \to Y \bar q_{Li} (s_{\b} \s^{\m}_{ij} \ph_{\m}, c_{\b} \s^{\m}_{ij} \ph_{\m}) \column{u_{Rj}}{d_{Rj}} \\
& \to \bar q_{Li} \, Y  \s^{\m}_{ij} U_{\m 0} \, \Ph_{0} q_{Rj} .
\end{align}
If we identify $\Ph_{0}$ as the Higgs boson, 
the Yukawa matrices found to be $Y_{q} = Y \s^{\m}_{ij} U_{\m 0}.$
Or, in components, 
\begin{align}
Y_{q}= Y \s^{\m} U_{\m 0} = Y 
\begin{pmatrix}
U_{00} + U_{30} & U_{10} - i U_{20} \\
U_{10} + i U_{20} & U_{00} - U_{30}
\end{pmatrix} . 
\end{align}
Therefore, when the heavy Higgs fields $\Ph_{i}$ are integrated out, 
the massless Higgs field $\Ph_{0}$ will have nontrivial flavor structures. 
In the following, we consider  
a concrete mass matrix which induces realistic flavor structures.

%%%%%%%%%%%%%%%%%%%%%%%%
\subsection{$U(2)_{L} \times U(2)_{R}$ breaking mass terms}
%%%%%%%%%%%%%%%%%%%%%%%%

The general analysis of the Higgs potential %4 $\times$ 4 matrix
is troublesome because there are so many free parameters. 
Then, we tentatively assume several interactions for 
a desirable mass matrix that induces 
proper eigenstates and mixing. 

The flavor symmetry is partially broken 
by the vev of the scalar $\h_{ij}$:
\begin{align}
\vev{\h_{ij}} = \diag{V}{0}. 
\end{align}
The spontaneous symmetry breaking (SSB) is transmitted to the yukawa coupling 
by interactions between $\h$ and $\ph$ fields. 

Here, we assume the following interactions
\begin{align}
\Lg_{i} = \sum_{a,b} a_{1} \tr [ \ph^{ab} \ph^{ba \dg} \h \h^{\dg}] + a_{2} \tr [ \ph^{ab \dg} \ph^{ba} \h \h^{\dg}] ,
\label{mass1}
\end{align}
with coupling constants $a_{1,2}$. 
The trace is summed over flavor indices $i, j$.  
According to the SSB, these interactions induce flavor violating mass of $\ph_{ij}^{ab}$ 
(hereafter gauge indices $a,b$ are abbreviated)
\begin{align}
\Lg_{i} \to V^{2} [ a_{1} (|\ph_{11}|^{2} + |\ph_{12}|^{2} ) + a_{2} (|\ph_{11}|^{2} + |\ph_{21}|^{2} )  ]. 
\end{align}
Note that the $\ph_{22}$ field remain massless (at tree level). 

This SSB induces several massless NGBs. 
In order to produce finite mass of these NGBs, 
we introduce the mass term that violates $U(2)_{L} \times U(2)_{R}$ explicitly:
\begin{align}
\Lg_{e} &= \sum_{a,b} \tr[\ph^{ab} \ph^{ba \dg} \offdiag{m^{2}}{m^{2}}] + \tr[\ph^{ab \dg} \ph^{ba} \offdiag{m^{2}}{m^{2}}] \\
& = m^{2}
[ \ph_{11} \ph_{21}^{*} + \ph_{12} \ph_{22}^{*} + \ph_{21} \ph_{11}^{*} + \ph_{22} \ph_{12}^{*} ]
+ m^{2} [ \ph_{11}^{*} \ph_{12} + \ph_{21}^{*} \ph_{22} + \ph_{12}^{*} \ph_{11} + \ph_{22}^{*} \ph_{21} ] .
\label{mass2}
\end{align}
Here, $m \ll V$ is a mass parameter. 
By these terms, the $U(2)_{L} \times U(2)_{R}$ flavor symmetry is broken to $U(1)_{V}$, 
and flavor mixing is introduced.

From Eqs.~(\ref{mass1}) and (\ref{mass2}), 
the mass matrix of $\ph_{ij}$ is found to be
\begin{align}
\ph^{*} \M \ph \equiv
\RRow{\ph_{11}}{\ph_{12}}{\ph_{21}}{\ph_{22}}^{*}
\begin{pmatrix}
M_{1}^{2} + M_{2}^{2} & m^{2} & m^{2} & 0 \\
m^{2} & M_{1}^{2} & 0 & m^{2} \\
m^{2} & 0 & M_{2}^{2} & m^{2} \\
0 & m^{2} & m^{2} & 0 \\
\end{pmatrix}
\CColumn{\ph_{11}}{\ph_{12}}{\ph_{21}}{\ph_{22}} ,
\label{mass3}
\end{align}
with $M_{1,2}^{2} \equiv a_{1,2} V^{2}$.
This mass matrix is perturbatively diagonalized as 
$U_{1}^{\dg} U_{2}^{\dg} \M U_{2} U_{1} = \M^{\rm diag}$, $\M^{\rm diag} = {\rm diag} ({M_{1}^{2} + M_{2}^{2}}, \, {M_{1}^{2}}, \, {M_{2}^{2}}, \, {- {m^{4} \over M_{1}^{2} } - {m^{4} \over M_{2}^{2}} } )$, where
\begin{align}
U_{1} = 
\begin{pmatrix}
 1 & 0 & - \l_{R} & 0 \\
 0 & 1 & 0 & - \l_{R} \\
\l_{R} & 0 & 1 & 0 \\
 0 & \l_{R} & 0 & 1 \\
\end{pmatrix} ,  ~~~~
U_{2} = 
\begin{pmatrix}
 1 & - \l_{L} & 0 & 0 \\
\l_{L} & 1 & 0 & 0 \\
 0 & 0 & 1 & - \l_{L} \\
 0 & 0 & \l_{L} & 1 \\
\end{pmatrix} ,
\end{align}
with $\l_{R,L} = m^{2} / M_{1,2}^{2}$.
The mass eigenstate $\Ph$ is defined as
 $\Ph \equiv U_{1}^{\dg} U_{2}^{\dg} \ph_{ij}$.
The fourth component of $\Ph$ is almost massless. 
Then the original field $\ph_{ij}$ is rewritten by $\Ph$ as
\begin{align}
\CColumn{\ph_{11}}{\ph_{12}}{\ph_{21}}{\ph_{22}}
= 
U_{2} U_{1}
\CColumn{\Ph_{1}}{\Ph_{2}}{\Ph_{3}}{\Ph_{0}}
= 
\CColumn{\frac{m^2}{M_{2}^{2} M_{1}^{2}} \Ph_{0} + \cdots }
{- \frac{ m^{2}}{M_{1}^{2}} \Ph_{0} + \cdots  }
{-  \frac{ m^{2}}{M_{2}^{2}} \Ph_{0} + \cdots }
{ \Ph_{0} + \cdots } . 
\end{align}
When the heavy fields $\Ph_{1,2,3}$ are integrated out from the low-energy theory, 
we obtain the hierarchical (lopsided) Yukawa matrix
\begin{align}
\Lg_{Y} = Y \bar q_{Li}^{a} \ph_{ij}^{ab} q^{b}_{Rj} + h.c. 
\to Y_{ij} \bar q_{Li}^{a} \Ph_{0}^{ab} q^{b}_{Rj} + h.c., 
\end{align}
where 
\begin{align}
Y_{ij} \simeq Y 
\begin{pmatrix}
\l_{L} \l_{R} & -\l_{R} \\
-\l_{L} & 1
\end{pmatrix} .
\end{align}

This example is rather artificial, 
nevertheless, 
it shows nontrivial flavor structure from mixing of flavored Higgs doublet. 
The Yukawa matrix can be interpreted as a result of  higher dimensional operator 
\cite{Anderson:1993fe}
\begin{align}
\begin{pmatrix}
0 & \l \\ \l & 1 
\end{pmatrix}
\sim 
\begin{pmatrix}
0 & {\vev{\th} \over \vev{\D}} \\
{\vev{\th} \over \vev{\D}} & 1
\end{pmatrix},
~~ \Gets ~~
\Lg =  {\th \over \vev{\D}} \tilde H (\bar q_{L1} u_{R2} + \bar q_{L2} u_{R1}) + h.c. \, .
\end{align}
However, this flavor structure is in principle 
originated from renormalizable Higgs potential. 
Meanwhile, in this example, mass of the 
lightest boson becomes negative $M_{0} = - {m^{4} \over M_{1}^{2}} - {m^{4} \over M_{2}^{2}}$ and 
triggers the electroweak symmetry breaking
(bosonic seesaw mechanism) \cite{Calmet:2002rf, Haba:2015lka,Haba:2015yfa}. 
Interplay between these ideas also seems to be interesting possibility.

%%%%%%%%%%%%%%%%%%%%%%%%%%%%%%%
\subsection{Fine-tunings of Higgs bosons } 
%%%%%%%%%%%%%%%%%%%%%%%%%%%%%%%

The lightest Higgs field in this example is not protected 
any symmetry and then the mass parameters 
should be fine-tuned when quantum corrections are included. 
On the other hand, 
the flavor symmetry is spontaneously broken and NGB emerges. 
Then, it is natural to consider 
whether the flavored Higgs bosons can be some NGBs or not. 
This kind of theory contains 
composite Higgs \cite{Georgi:1984af,Kaplan:1983fs,Agashe:2004rs}, Little Higgs \cite{ArkaniHamed:2002qy,ArkaniHamed:2002qx}, twin Higgs \cite{Chacko:2005pe}, 
GIFT mechanism with supersymmetry \cite{Inoue:1985cw}, and so on. 
If the Higgs boson is a gauged NGB, 
the mass parameters Eqs.~(\ref{mass1}) and (\ref{mass2}) (or Eq.~(\ref{mass3})) can be treated 
as breaking of original symmetry, like in the chiral perturbation theory \cite{Pich:1998xt}. 

In a theory with gauged NGBs, 
usually gauge interactions violate the global symmetry 
which protect the NGBs massless. 
Then these theories require some TeV-scale partner fields 
to restrain quadratic divergences. 
However, direct and indirect searches by LHC and several flavor experiments 
severely restrict such TeV-scale partners. 
Therefore, in this idea, it is difficult to avoid fine-tunings 
unless an essentially new mass protection mechanism of scalar fields. 

%%%%%%%%%%%
\section{Conclusions}
%%%%%%%%%%%

In this paper, we suggest a simple model 
which induces realistic flavor structure from 
mixing of flavored Higgs doublets. 
The idea is based on the decoupling limit. 
In a model with many Higgs doublets, 
 the mass eigenstates of scalars are linear combinations of Higgs doublets. 
If the mass matrix of Higgs fields has only one massless mode, 
and if the linear combination has flavor dependence, 
the induced Yukawa coupling will have nontrivial flavor structure.
We suggest a mass matrix of flavored Higgs fields in a $U(2)_{L} \times U(2)_{R}$ toy model. 

An advantage of this model 
is that all of the elements in Yukawa matrix 
can be determined from renormalizable Higgs potential. 
This feature is substantially different from Froggatt--Nielsen mechanism  
with many undetermined $\mathcal O(1)$ couplings. 
A drawbacks are arbitrariness of mass parameters 
and fine-tunings. The heavy flavored Higgs bosons have ``natural mass''. 
By contrast, the mass of the lightest SM Higgs boson should be fine-tuned. 
If the flavored Higgs fields are gauged NGBs, 
such as in composite Higgs \cite{Georgi:1984af,Kaplan:1983fs,Agashe:2004rs}, Little Higgs \cite{ArkaniHamed:2002qy,ArkaniHamed:2002qx}, twin Higgs \cite{Chacko:2005pe}, 
GIFT mechanism with supersymmetry \cite{Inoue:1985cw}, 
these theories require TeV-scale partner fields that have strong tension 
with current experiments. 
Therefore, in this idea, it is difficult to avoid fine-tunings of the lightest Higgs mass.

Meanwhile, if the determinant of the mass matrix $\det \M < 0$ holds, mass of the lighter boson becomes negative and 
triggers the electroweak symmetry breaking
(bosonic seesaw mechanism). 
Interplay between these ideas also seems to be interesting possibility.

%%%%%%%%%%%%%%
\section*{Acknowledgement}
%%%%%%%%%%%%%%

This study is financially supported by the Iwanami Fujukai Foundation.

%\bibliographystyle{bib/h-physrev50}
%\bibliography{bib/Higgsmixing}

\end{document}